\documentclass[12pt,twocolumn]{emulateapj}

\shorttitle{Escaping CR Electron Spectrum}
\shortauthors{Kawanaka et al.}

\begin{document}

\title{TeV Electron Spectrum for Probing Cosmic-Ray Escape from a Supernova Remnant}
\author{Norita Kawanaka\altaffilmark{1}, Kunihito Ioka\altaffilmark{1}, Yutaka Ohira\altaffilmark{1} \& Kazumi Kashiyama\altaffilmark{2}}
\altaffiltext{1}{Theory Center,
Institute of Particle and Nuclear Studies,
KEK (High Energy Accelerator Research Organization),
1-1 Oho, Tsukuba 305-0801, Japan} 
\altaffiltext{2}{Department of Physics, Kyoto University, Kyoto 606-8502, Japan}
\email{norita.kawanaka@kek.jp}

\begin{abstract}
One of the most essential but uncertain processes for producing
cosmic-rays (CRs) and their spectra
is how accelerated particles escape into the interstellar space.  We
propose that the CR electron
spectra at $\gtrsim {\rm TeV}$ energy can provide a direct probe of
the CR escape complementary to the CR nuclei and gamma-rays.  We calculate the electron spectra from
a young pulsar embedded in the supernova remnant (SNR), like Vela, taking into account the energy-dependent CR escape.
  SNRs would accelerate and hence confine particles with energy
 up to $10^{15.5}{\rm eV}$.  Only energetic particles can escape first, while the lower energy particles are confined and released later.
Then the
observed electron spectrum
should have a low energy cutoff whose position marks the age of the
pulsar/SNR.  The low energy
cutoff is observable in the $\gtrsim {\rm TeV}$ energy window, where
other contaminating sources are
expected to be few due to the fast cooling of electrons.
The spectrum looks similar to a dark matter annihilation line
if the low energy cutoff is close 
to the high energy intrinsic or cooling break.
The future experiments such as
CALET and CTA are capable of directly detecting the CR escape features toward revealing
the origin of CRs.
\end{abstract}
\keywords{acceleration of particles -- cosmic rays -- pulsars:general -- supernova remnants}

\section{Introduction}

The origin of cosmic-rays (CRs) is a long standing problem since its discovery.  The number spectrum of the nuclear component of CRs can be fitted with a broken power-law: $N(\varepsilon) \propto \varepsilon^{-s}$ with $s\simeq 2.7$ below $\varepsilon\simeq 10^{15.5}{\rm eV}$ (the "knee" energy).  The charged CR particles are considered to propagate diffusively with an energy-dependent manner (the higher energy particles can diffuse faster), so the intrinsic CR spectrum at the source should be harder than that observed at the Earth.  Although there are some ambiguities in the energy dependence on the diffusion coefficient, the source spectral index is considered to be around $s\simeq 2.2-2.4$ (e.g. Strong \& Moskalenko 1998).   

It is often argued that the CRs with energy smaller than the knee energy (or even up to $10^{18}{\rm eV}$) are originated in our Galaxy.  The most widely accepted paradigm for the galactic CR production process is the diffusive shock acceleration (DSA) at supernova remnants (SNRs).  The theory of DSA (for reviews see Blandford \& Eichler 1987; Malkov \& Drury 2001) can naturally derive the power-law spectrum of particles accelerated in the SNR shock.  In fact, H.E.S.S. detected TeV gamma-ray emissions from the shell of young SNRs (Aharonian et al. 2004, 2005), and Fermi and AGILE have detected GeV gamma-ray emissions from middle-aged SNRs interacting with nearby molecular clouds, which are likely to be generated via hadronic process, i.e. inelastic collisions between CR protons accelerated at SNRs and ambient protons (Abdo et al. 2009b, 2010a, b, c; Tavani et al. 2010; Giuliani et al. 2010).  However, in the conventional DSA theory the predicted spectral index of CRs accelerated at the shock with a large Mach number is $s=2$ and harder than that expected from the observations.  Moreover, the non-linear theories of DSA predict a harder source spectrum $s\lesssim 2.0$, which seems to make the contradiction between theories and observations even worse.

It has been proposed that these observational facts can be interpreted by the model of the energy-dependent CR escape from the SNR (Ptuskin \& Zirakashvili 2005; Drury et al. 2009; Reville et al. 2009; Caprioli et al. 2009; Gabici et al. 2009; Ohira et al. 2010a; Fujita et al. 2010; Casanova et al. 2010).  In the escape model it is assumed that the most energetic particles leave the SNR at the beginning of the Sedov phase and, as the shock slows down and the magnetic field decays, the lower energy particles, which have been confined around the shock front by the magnetic field, can leave the shock gradually.  This process can largely affect the CR spectrum detected at the Earth (Ohira et al. 2010a; Caprioli et al. 2010).  In fact, the spectral index of the CR escaping the SNR $s_{\rm esc}$ is different from that of the CR inside the SNR $s$,
\begin{eqnarray}
s_{\rm esc}=s+\frac{\beta}{\alpha},
\end{eqnarray}
where the escape energy $\varepsilon_{\rm esc}$ and the normalization factor of the CR production rate are assumed to be proportional to $t^{-\alpha}$ and $t^{\beta}$, respectively.  For example, in the phenomenological model by Gabici et al. (2009) $\alpha\sim 2.6$ (see also Ohira et al. 2010a).
Therefore the observed spectral index can get softer than the conventional value ($s\simeq 2$) if the normalization factor of the spectrum gets larger with time (i.e. $\beta>0$; Ohira et al. 2010a), so we may interpret the observed CR spectrum in the context of DSA with the escape model.  Moreover, recent gamma-ray observations of middle-aged SNRs interacting with molecular clouds by Fermi and AGILE show that those spectra are fitted by a broken power-law with a break energy of $\sim 1-10{\rm GeV}$ (see the references shown above), and they can be interpreted by the energy-dependent CR escape from an SNR (Aharonian \& Atoyan 1996; Gabici et al. 2009; Ohira et al. 2010b)\footnote{Recently some authors have tried to explain this spectral break by considering the other processes such as the propagation of the SNR shock in the molecular clouds (Uchiyama et al. 2010), or the two-step acceleration of cosmic-ray particles by the shocks generated by the turbulence behind the SNR shock (Inoue et al. 2010).}.  However, these gamma-rays are the secondary emissions of CRs from an SNR, and so these observations are only the indirect evidences of the CR escape scenario.

On the other hand, the direct observations of CR electrons/positrons have been also greatly advanced.  The PAMELA satellite discovered the excess of the CR positron fraction (Adriani et al. 2009) and, as for the flux of CR electrons plus positrons the experiments such as ATIC/PPB-BETS (Chang et al. 2008; Torii et al. 2008b), Fermi (Abdo et al. 2009a; Ackermann et al. 2010) and H.E.S.S. (Aharonian et al. 2008, 2009) have revealed an excess from the conventional model (see also Meyer et al. 2010 for the energy calibration between Fermi and H.E.S.S.).  These results suggest that there are some additional electron/positron sources.  Possible candidates include astrophysical sources such as pulsars (Shen 1970; Atoyan et al. 1995; Chi et al. 1996; Zhang \& Cheng 2001; Grimani 2007; Kobayashi et al. 2004; B\"{u}esching et al. 2008; Hooper et al. 2009; Yuksel et al. 2009; Profumo 2008; Malyshev et al. 2009; Grasso et al. 2009; Kawanaka et al. 2010; Heyl et al. 2010; Blasi \& Amato 2010), supernova remnants (Shaviv et al. 2009; Blasi 2009; Blasi \& Serpico 2009; Fujita et al. 2009; Hu et al. 2009; Biermann et al. 2009; Mertsch \& Sarkar 2009; Ahlers et al. 2009), gamma-ray bursts (GRB; Ioka 2010; Kistler \& Yuksel 2009), microquasars (Heinz \& Sunyaev 2002) and white dwarf pulsars (Kashiyama et al. 2010).  Dark matter annihilations/decays (e.g. Hooper 2009) and the propagation effect (Delahaye et al. 2008; Cowsik \& Burch 2009; Stawarz et al. 2010) are also the possible processes for making the CR electron/positron excess (for the comprehensive review, see Fan et al. 2010).

In the near future within a few years, Alpha Magnetic Spectrometer - 02 (AMS-02) experiment will measure the positron fraction up to $\sim 1{\rm TeV}$ (Beischer et al. 2009; Pato et al. 2010; Pochon 2010) and CALorimetric Electron Telescope (CALET) experiment will explore the electron spectrum\footnote{Hereafter we express the spectrum of electrons plus positrons as just "electron spectrum".} up to $\sim 10{\rm TeV}$ with an energy resolution better than a few percent (Torii et al. 2008a).  In addition, the future Cherenkov Telescope Array (CTA) will be able to measure the CR electron spectrum up to $\sim 15{\rm TeV}$ (CTA consortium 2010).
These experiments will open the window to the CR astrophysics with the TeV electron and positron components.  Especially, as Kobayashi et al. (2004) have pointed out, in the TeV energy band a few nearby sources may leave spectral signatures
 and we will be able to see a spectral shape of CR electrons/positrons from a single source, while in the lower energy band we can see only the superposed spectrum from multiple sources.

In this paper, we suggest a possibility that the precise measurement of the CR electron spectrum in the very high energy band ($\gtrsim 1-10{\rm TeV}$, expected with CALET and CTA) could directly prove the energy-dependent CR escape if those electrons/positrons are generated by a pulsar embedded in an SNR.  This is quite a natural situation to be realized because a pulsar should be generated in the center of a core-collapse supernova and CR electrons/positrons are considered to be generated in the pulsar wind nebula formed inside the SNR.  If the CR escape really occurs in a young nearby pulsar/SNR system, the electron spectrum from it will show a unique feature as explained in the following and, when it is observed, that will be the first direct evidence of the CR escape scenario.

\section{A Simple Example}

To illustrate our main idea, we show a clear example of the effect of the energy-dependent CR escape in the direct electron observations.
  In the calculations of CR electrons/positrons, we usually assume that they are injected into the interstellar matter (ISM) from a pulsar with a spectral form of
\begin{eqnarray}
Q_e(\varepsilon_e) =Q_0 \varepsilon_e^{-\alpha},
\end{eqnarray}
where $\varepsilon_e$ is the energy of electrons/positrons.  The observed electron spectrum $f(\varepsilon_e,r,t)$ can be obtained by solving the
 diffusion equation
\begin{eqnarray}
\frac{\partial }{\partial t}f= D(\varepsilon_e)\nabla ^2 f+\frac{\partial }{\partial \varepsilon_e}[P(\varepsilon_e)f]+Q_e(\varepsilon_e,r,t), \label{diffeq}
\end{eqnarray}
where $D(\varepsilon_e)=D_0(1+\varepsilon_e/3{\rm GeV})^{\delta}$ is the diffusion coefficient and $P(\varepsilon_e)$ is the energy loss rate.  Here we adopt $D_0=5.8\times 10^{28}{\rm cm}^2{\rm s}^{-1}$, $\delta=1/3$ that is consistent with the boron to carbon ratio according to the latest GALPROP code, and $P(\varepsilon_e)=-b\varepsilon_e^2$ with $b=10^{-16}{\rm GeV}^{-1}{\rm s}^{-1}$ which includes the energy loss via synchrotron emission and inverse Compton scatterings (with Thomson approximation).  Then, if electrons/positrons are injected from a point-like source instantaneously (i.e. $Q_e(\varepsilon_e,r,t)\propto \delta(t)\delta(r)$), the observed spectrum is simply written as

\begin{eqnarray}
f\sim \frac{Q_0 \varepsilon_e^{-\alpha}}{\pi^{3/2} d_{\rm diff}^3}
(1-b t \varepsilon_e)^{\alpha-2}
e^{-(r/d_{\rm diff})^2},
\label{eq:fsol}
\end{eqnarray}
where $d_{\rm diff}\sim \sqrt{4D(\varepsilon_e)t}$ is the diffusion length of CR electrons/positrons.  This spectrum is roughly proportional to $\varepsilon_e^{-\alpha-3\delta/2}\exp(-r^2/(4D(\varepsilon_e)t))$ up to the sharp cutoff at $\varepsilon_e\sim 1/(bt)$, and exponentially damps beyond the diffusion length $d_{\rm diff}$ (Atoyan et al. 1995; Ioka 2010).

However, if the energy-dependent escape of CR particles from the shock is taken into account, the electron spectrum would have a
sharp cutoff in the low energy side because lower energy CRs cannot escape into the ISM.
  The energy of particles which are marginally capable of escaping to the ISM $\varepsilon_{\rm esc}$ is generally determined by the confinement condition of CR particles, i.e. the equality between
 the diffusion length of particles and the characteristic size of the system:
\begin{eqnarray}
l_{\rm diff}=\frac{D_{\rm sh}(\varepsilon_{\rm esc})}{u_{\rm sh}}\sim R_{\rm sh},
\end{eqnarray}
where $D_{\rm sh}$, $u_{\rm sh}$ and $R_{\rm sh}$ are the diffusion coefficient around the supernova remnant shock, the shock velocity, and
 the size of the system, respectively.  Then the energy $\varepsilon_{\rm esc}$ is generally a function of time $t$ due to the evolution of $D_{\rm sh}$, $R_{\rm sh}$ and $u_{\rm sh}$.

Fig.~1 shows the electron spectrum assuming that
only electrons and positrons above the escape energy $\varepsilon_{\rm esc}$
are injected instantaneously.  The energy of electrons/positrons which is initially $\varepsilon_{\rm esc}$ becomes $\varepsilon_{\rm esc}/(1+b\varepsilon_{\rm esc}t_{\rm age})$ as a result of radiative cooling (where $t_{\rm age}$ is the time since the emission), and we can clearly see a sharp cutoff of the spectrum at that energy, which is the effect of energy-dependent escape of CR particles.  If such a spectrum is confirmed by the future experiments, it would be the strong support for the CR escape scenario of the particle acceleration process at the SNR, and we may also get the information of $\varepsilon_{\rm esc}$ at the time of the source age
$t_{\rm age}$.

\begin{figure}

\plotone{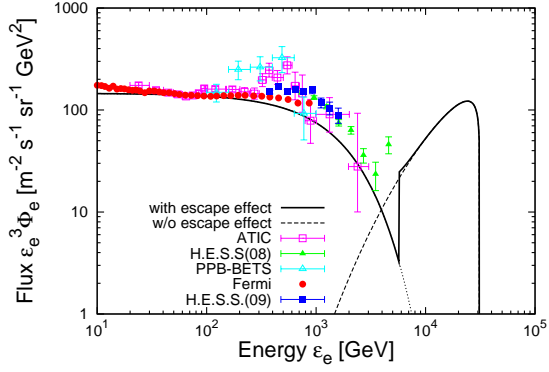}

\caption{The electron spectrum 
predicted from a transient source with the escape energy of $\varepsilon_{\rm esc}=7{\rm TeV}$ (solid line)
added with the background model (dotted line),
compared with the ATIC/PPB-BETS/H.E.S.S./Fermi data.  The spectrum without assuming the energy-dependent escape is also shown (dashed line). 
We assume that a source at $r=700{\rm pc}$
from the Earth produces $e^{\pm}$ pairs a time $t_{\rm age}=1.0\times 10^4{\rm years}$ ago with total energy $E_{e^+}=E_{e^-}=0.5\times 10^{49}{\rm erg}$
and spectral index $\alpha=2.4$.
}

\label{f1}

\end{figure}
\section{Model of CR Electron/Positron Escape}

Let us consider the more sophisticated model than discussed in the previous section.  We consider a pulsar emitting electrons and positrons embedded in the SNR (i.e. it has not been evacuated from the SNR by the natal kick).  A pulsar is considered to be an efficient $e^{\pm}$ factory, because its rotating magnetic field would produce
 a strong electric field around the pulsar and then a large number of $e^{\pm}$ pairs would be produced via electromagnetic cascades.  The created pairs would stream away by a centrifugal force as a pulsar wind that ends with a termination shock where the acceleration of electrons/positrons may occur.  Hereafter we assume the $e^{\pm}$ production rate per energy from a pulsar having a spectrum with a cutoff power-law shape:
\begin{eqnarray}
\dot{N}_{e,{\rm pr}}(\varepsilon_e,t)=Q_0 (t)\varepsilon_e^{-\alpha}\exp \left(-\frac{\varepsilon_e}{\varepsilon_{e,{\rm cut}}} \right),
\end{eqnarray}
where the high energy break is fixed as $\varepsilon_{e,{\rm cut}}=10{\rm TeV}$.  According to the gamma-ray observations of PWNe, the electrons/positrons accelerated up to the energy of 10-100TeV seem to exist in the nebulae (e.g., Aharonian et al. 2006), so this assumption is reasonable.
Here $Q_0(t)$ is given to satisfy  
\begin{equation}
L_{e,{\rm pr}}(t) = \int _{\varepsilon_{e,{\rm min}}}^{\infty } d\varepsilon_e \varepsilon_e \dot{N}_{e,{\rm pr}}(\varepsilon_e,t),
\end{equation}
where $L_{e,{\rm pr}}(t)$ is the electron/positron production luminosity and $\varepsilon_{e,{\rm min}}$ is set to be $1{\rm GeV}$.  We assume that this luminosity is proportional to the spin-down luminosity of the pulsar:
\begin{eqnarray}
L_{e,{\rm pr}}(t)\propto \frac{1}{(1+t/\tau_0)^2},
\end{eqnarray}
where $\tau_0\sim 10^{2-4}{\rm year}$ is the spin-down timescale, which is related to the surface magnetic field of the pulsar (Shapiro \& Teukolsky 1983).

In order to evaluate the CR spectrum in the escape scenario, we should assume the time evolution of $\varepsilon_{\rm esc}$.  In this study we adopt two models for the functional form of $\varepsilon_{\rm esc}$.  In the first model, we assume for $\varepsilon_{\rm esc}(t)$ a power-law behavior, and determine its normalization and power-law index according to the hypothesis that SNRs are responsible for the observed CRs with the energy from $\sim 1{\rm GeV}$ up to the knee energy ($\sim 10^{15.5}{\rm eV}$).  Then $\varepsilon_{\rm esc}(t)$ should reach the knee energy at the end of the free expansion phase (i.e. the beginning of the Sedov phase; $t_{\rm Sedov}$) and should decrease down to 1GeV at $t\simeq 10^{5/2}t_{\rm Sedov}$ (i.e. the end of the SNR expansion; Gabici et al. 2009; Ohira et al. 2010a):
\begin{eqnarray}
\varepsilon_{\rm esc}(t)=10^{6.5}{\rm GeV}\times \left(\frac{t}{t_{\rm Sedov}} \right)^{-2.6}.
\end{eqnarray}

As the second model for the evolution of $\varepsilon_{\rm esc}$ we adopt the one discussed by Ptuskin \& Zirakashvili (2005), which takes into account the modification of a shock structure due to the CR pressure, as well as the non-linear dissipation of magnetic turbulence.  They solve the steady-state equation which determines the energy density of the magnetohydrodynamic turbulence $W$:
\begin{eqnarray}
u\nabla W(k)=2\left( \Gamma_{\rm cr}(k)-\Gamma_{\rm l}(k)-\Gamma_{\rm nl}(k) \right) W(k),
\end{eqnarray}
where $u$ is the flow velocity (here it is equal to the shock velocity $u_{\rm sh}$), $k$ is the wave number of the turbulence, and $\Gamma_{\rm cr}$, $\Gamma_{\rm l}$ and $\Gamma_{\rm nl}$ are the
 wave growth rate at the shock due to the CR streaming instability, the damping rate of waves in the background plasma due to the ion-neutral and electron-ion collisions (linear damping), and due to the wave-wave interactions (non-linear damping), respectively.
   The mathematical expressions for these functions are shown in Ptuskin \& Zirakashvili (2005), and by solving this equation (while the term of linear damping is neglected), we obtain the threshold particle energy for escape as a function of the shock velocity $u_{\rm sh}$.  Since
 we know the time dependence of the shock radius and the shock velocity in the Sedov phase, we can
 derive $\varepsilon_{\rm esc}(t)$ as a function of time.  When the age of the SNR is younger than $\lesssim 10^5{\rm years}$, the evolution of $\varepsilon_{\rm esc}(t)$ in this model can be approximated as
\begin{eqnarray}
\varepsilon_{\rm esc}(t)\simeq 10^{5.5}{\rm GeV}\times \left(\frac{t}{t_{\rm Sedov}} \right)^{-1.0}.
\end{eqnarray}

  Fig. 2 shows the evolutions of
 $\varepsilon_{\rm esc}(t)$ in two models described above.  Once we fix the time dependence of $\varepsilon_{\rm esc}$, we can describe the number luminosity per energy of escaping electrons and positrons as
\begin{eqnarray}
\dot{N}_{e,{\rm esc},1}(\varepsilon_e,t)=\dot{N}_{e,{\rm pr}}(\varepsilon_e,t) \Theta(\varepsilon_e-\varepsilon_{\rm esc}),
\end{eqnarray}
where $\Theta (x)$ is the step function\footnote{Strictly speaking, the spectrum of the escape flux has a finite width around the escape energy $\varepsilon_{\rm esc}$.  However in the usual case the step function is a good approximation (Ptuskin \& Zirakashvili 2005; Caprioli et al. 2009).}.

\begin{figure}

\plotone{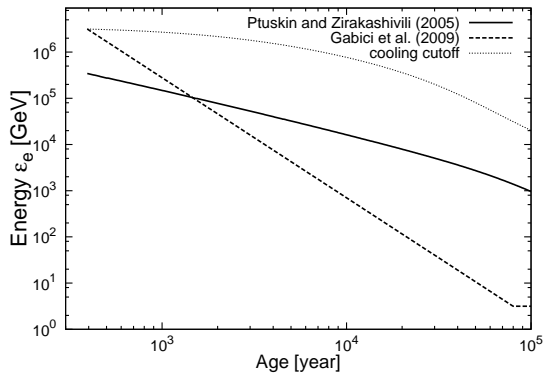}

\caption{The threshold energy for escape from the SNR shock $\varepsilon_{\rm esc}$ in the Sedov phase as a function of time.  The thick solid line and the thin solid line correspond to the model by Ptuskin \& Zirakashvili (2005) and Gabici et al. (2009), respectively.  The cutoff which appears in the spectrum due to the energy loss (including the KN effect) during the propagation is also shown (dashed line).
}

\label{f2}

\end{figure}

There is another flux component which should be taken into account.  The electrons/positrons which have the energy lower than $\varepsilon_{\rm esc}(t)$ are confined in the SNR and lose their energy adiabatically (Ptuskin \& Zirakashvili 2005).  As $\varepsilon_{\rm esc}(t)$ decreases with time,
 some of the confined and adiabatically cooled electrons/positrons can escape the shock surface when their energy becomes greater than $\varepsilon_{\rm esc}(t)$.  First, the CR electron/positron number per energy confined in the SNR can be written as
\begin{eqnarray}
N_{e,{\rm conf}}(\varepsilon_e,t)&=&\int_{t_{\rm Sedov}}^{t} dt^{\prime} \dot{N}_{e, {\rm pr}}(\varepsilon_e^{\prime},t^{\prime}) \frac{d\varepsilon_e^{\prime}}{d\varepsilon_e} \nonumber \\
&&\times \Theta \left( \varepsilon_{\rm esc}(t^{\prime})-\varepsilon_e^{\prime}  \right),
\end{eqnarray}
where $\varepsilon_e^{\prime}$ is the energy of the electrons/positrons at the time $t^{\prime}$, and it is $\varepsilon_e$ at the time $t$.
  
The adiabatic loss is determined by the expansion law of the SNR,
\begin{eqnarray}
\frac{d\varepsilon_e}{dt}=-\frac{\dot{R}_{\rm SNR}}{R_{\rm SNR}}\varepsilon_e,
\end{eqnarray}
where $R_{\rm SNR}$ and $\dot{R}_{\rm SNR}$ are the radius of the SNR shell and its expansion velocity, respectively.  In the following we assume that the SNR is in the Sedov phase, in which the time dependence of $R_{\rm SNR}$ is expected to be proportional to $t^{2/5}$.
Then the energy of electrons/positrons confined in the SNR would evolve as
\begin{eqnarray}
\varepsilon_e(t)=\varepsilon_e^{\prime} \left( \frac{t^{\prime}}{t} \right)^{2/5},
\end{eqnarray}
and therefore, the distribution function of confined electrons/positrons can be evaluated as
\begin{eqnarray}
N_{e,{\rm conf}}(\varepsilon_e,t)&=&\int_{t_{\rm Sedov}}^{t} dt^{\prime} Q_0 (t^{\prime})\varepsilon_e^{-\alpha}\left( \frac{t}{t^{\prime}} \right)^{2/5(1-\alpha)} \nonumber \\
&\times &\exp \left( -\frac{\varepsilon_e (t/t^{\prime})^{2/5}}{\varepsilon_{e,{\rm cut}}} \right) \nonumber \\
&\times & \Theta \left( \varepsilon_{\rm esc}(t^{\prime})-\varepsilon_e\left( \frac{t}{t^{\prime}} \right)^{2/5} \right).
\end{eqnarray}
If the decreasing rate of $\varepsilon_{\rm esc}$ is faster than the adiabatic cooling rate, $-(2/5)\varepsilon_{\rm esc}/t$, then
a part of confined electrons/positrons can escape the SNR shock.  By using the same logic in deriving Eq. (21) of Ptuskin \& Zirakashvili (2005) in the case of expanding media, we can evaluate the spectrum of such particles as
\begin{eqnarray}
\dot{N}_{e,{\rm esc},2}(\varepsilon_e,t)&=&-\delta(\varepsilon_e-\varepsilon_{\rm esc}(t))
 \left(\frac{\partial \varepsilon_{\rm esc}}{\partial t}+\frac{2\varepsilon_e}{5t}\right)\nonumber \\
&&\times N_{e,{\rm conf}}(\varepsilon_e,t). \label{escape2}
\end{eqnarray}

Hereafter we consider the spectrum of escaping electrons/positrons number luminosity per energy as the sum of above two components:
\begin{eqnarray}
\dot{N}_{e,{\rm esc}}(\varepsilon_e)=\dot{N}_{e,{\rm esc},1}(\varepsilon_e )+\dot{N}_{e,{\rm esc},2}(\varepsilon_e ).
\end{eqnarray}

We neglect the radiative energy loss of electrons/positrons during the confinement for simplicity.  As we have shown in Eq. (\ref{escape2}), the electron flux of the second component $\dot{N}_{e,{\rm esc},2}(\varepsilon_e,t)$ is determined by the difference between the decline rate of the escape energy $\varepsilon_{\rm esc}$ and the energy loss rate of electrons/positrons.  In the cases shown in this study, the decline rate of the escape energy is $\sim \alpha \varepsilon_e/t\sim 3\times 10^{-9}\left( \varepsilon_e/1{\rm TeV}\right) \left(t/10^4{\rm yr}\right) ^{-1}{\rm GeV}~{\rm sec}^{-1}$ ($\alpha\simeq 1-2.6$, depending on the model) , while the adiabatic cooling rate and the radiative cooling rate is $\sim 1.2\times 10^{-9}\left(\varepsilon_e/1{\rm TeV}\right) \left(t/10^4{\rm yr}\right) ^{-1}{\rm GeV}~{\rm sec}^{-1}$, $\sim b\varepsilon_e^2 \sim 10^{-10}\left(\varepsilon_e/1{\rm TeV}\right) ^2 {\rm GeV}~{\rm sec}^{-1}$, respectively, where for the latter we take the cooling rate for the interstellar space (see Sec. 2).  Therefore, even if we take into account the radiative cooling the flux of the electrons/positrons which have once been confined in the SNR does not change its order from our calculation.  Moreover, in the interstellar space the diffusion timescale for TeV electrons can be estimated as
\begin{eqnarray}
t_{\rm diff}&\sim &\frac{r^2}{4D(\varepsilon_e )} \nonumber \\
&\sim &1.4\times 10^4{\rm yr}\left( \frac{\varepsilon_e}{3{\rm TeV}} \right) ^{-1/3}\left( \frac{r}{300{\rm pc}}\right) ^2,
\end{eqnarray}
and therefore the energy loss of TeV electrons during the propagation is at most
$\Delta \varepsilon_e/\varepsilon_e\sim 1-(1+bt_{\rm diff}\varepsilon_e)^{-1}\sim 10\%$.  Then we can say that in this energy range
 both of the energy losses in the PWN and in the interstellar space are small.  However, if the magnetic field in the PWN/SNR is strongly amplified from the interstellar value, the energy loss rate due to the synchrotron emission may be faster than that due to the adiabatic expansion of the SNR, and even than that due to the decline rate of the escape energy.  In such case, the confined electrons/positrons cannot escape the SNR later and only the first component $\dot{N}_{e,{\rm esc},1}$ would be emitted and observed at the Earth.  Anyway, as either the strength of the magnetic field in the PWN/SNR nor its time evolution is generally uncertain, it is difficult to estimate the radiative cooling rate with a moderate strength of the magnetic field in a reliable way.  In the following, we investigate only the case that the radiative energy loss is not significant.

\section{Observed Electron Spectrum in the Escape Scenario}
Now that we have the time evolution of the escaping CR particle flux 
from the source, we can obtain the observed
electron spectrum by solving the propagation of CR electrons/positrons with 
the diffusion equation shown in Eq.(\ref{diffeq}).  Once we know the Green's function of this
 equation with respect to the time and position, $G(t,r,\varepsilon_e;\tau)$, we can obtain the observed electron spectrum as
\begin{eqnarray}
f(t,r,\varepsilon_e)&=&\int _{t_i} ^{t} G(t,r,\varepsilon_e;\tau)d\tau,
\end{eqnarray}
where $t_i$ is the time when the particle injection has started, which is assumed to be equal to $t_{\rm Sedov}$ in the following discussions.

The mathematical description of $G(t,r,\varepsilon_e;\tau)$ was derived by Atoyan et al. (1995),
\begin{eqnarray}
G(t,r,\varepsilon_e;t_0)=\frac{\dot{N}_{e,{\rm esc}} (\varepsilon_{e,0},t_0)P(\varepsilon_{e,0})}
{\pi ^{3/2} P(\varepsilon_e) d_{\rm diff}(\varepsilon_{e,0})^3}\exp \left(-\frac{r^2}{d_{\rm diff}(\varepsilon_{e,0})^2} \right),
\end{eqnarray}
where $\varepsilon_{e,0}$ is the energy of electrons/positrons at the time $t_0$ which are cooled down to $\varepsilon_e$ at the time $t$, and $d_{\rm diff}$ is the diffusion length given by
\begin{eqnarray}
d_{\rm diff}=2\left[\int_{\varepsilon_e}^{\varepsilon_{e,0}} \frac{D(x)dx}{P(x)}\right]^{-1/2},
\end{eqnarray}
as shown in Eqs.(10) and (11) in Atoyan et al. (1995).
 
 In deriving the energy loss rate $P(\varepsilon_e)$, we use the formulation shown by Moderski et al. (2005)
\begin{eqnarray}
P(\varepsilon_e)=\frac{4\sigma _T \varepsilon_e ^2}{3m_e^2 c^3}
\left[ \frac{B^2}{8\pi}+\int d\varepsilon_{\gamma} u_{\rm tot}(\varepsilon_{\gamma}) f_{\rm KN} 
\left( \frac{4\varepsilon_e \varepsilon_{\gamma}}{m_e^2 c^4} \right) \right] ,
\end{eqnarray}
where $\sigma_T$ is the Thomson cross section, $u_{\rm tot}(\varepsilon_{\gamma})d\varepsilon_{\gamma}$ is the energy density of interstellar photons with the energy between $\varepsilon_{\gamma}$ and $\varepsilon_{\gamma}+d\varepsilon_{\gamma}$ (including CMB, starlight and dust emission; Porter et al. 2008), and
 $B$ is the interstellar magnetic field which we here set as 1$\mu$G.  Here the function $f_{\rm KN}(x)$ is the correction factor to include the Klein-Nishina effect.  According to Moderski et al. (2005), this function can be expressed as
\begin{eqnarray}
f_{\rm KN}(\tilde{b})=\frac{9g(\tilde{b})}{\tilde{b}^3},
\end{eqnarray}
where $\tilde{b}=4\varepsilon_e \varepsilon_{\gamma}/(m_e c^2)^2$,
\begin{eqnarray}
g(\tilde{b})&=&\left(\frac{1}{2}\tilde{b}+6+\frac{6}{\tilde{b}}\right) \ln(1+\tilde{b}) \nonumber \\
 &&-\left(\frac{11}{12}\tilde{b}^3+6\tilde{b}^2+9\tilde{b}+4\right)\frac{1}{(1+\tilde{b})^2} \nonumber \\
&&-2+2{\rm Li}_2(-\tilde{b}),
\end{eqnarray}
and the function ${\rm Li}_2(z)$ is the dilogarithm
\begin{eqnarray}
{\rm Li}_2(z)=\int_z^0 \frac{\ln (1-t)dt}{t}.
\end{eqnarray}

According to the recent experiments, especially H.E.S.S. (Aharonian et al. 2008), the background CR electron/positron flux seems to
have a high energy dropping around a few TeV.  This dropping is quite naturally explained in the context of the astrophysical origin of CR electrons/positrons because
the number of the sources contributing to the TeV energy band is quite small according to the birth rate of SNe/pulsars in the vicinity of the Earth (Kobayashi et al. 2004; Kawanaka et al. 2010).  In fact, since the pulsars which contribute to the electron flux at the energy $\varepsilon_e$ should be younger
 than the cooling time of electrons/positrons $t_{\rm cool}\sim 1/(b\varepsilon_e)$ and should be located closer to the Earth than the diffusion length $d_{\rm diff}\sim 2\sqrt{K(\varepsilon_e)t_{\rm cool}}$, the number of the pulsars contributing to $\gtrsim {\rm TeV}$ band should be as small as
\begin{eqnarray}
N_{\rm PSR}(\varepsilon_e)\sim 6\left( \frac{\varepsilon_e}{{\rm TeV}} \right) ^{-5/3}
\left( \frac{R}{0.7\times 10^{-5} {\rm yr}^{-1}{\rm kpc}^{-2}} \right),
\end{eqnarray}
where $R$ is the local pulsar birth rate per unit surface area of our Galaxy.  If we can separate the contribution of a single young source from the observed electron spectrum, we can get the information of the CR injection into the ISM from that source.  For this reason, we especially pay attention to the TeV spectral features of CR electrons from a pulsar in the followings.

In Figs.~3 and 4 we show the time evolutions of CR electron spectrum from a nearby pulsar according to the models of the escape energy $\varepsilon_{\rm esc}(t)$ adopted in the previous section (see also Fig. 2).  It is clear that there exists a low energy cutoff in each
 spectrum corresponding to the value of $\varepsilon_{\rm esc}(t)$ at that time.  The spectral shapes generally depend on other parameters such as the high energy break of the intrinsic electron spectrum, the spectral index, the duration of electron/positron injection from a pulsar, and the total
 energy of CR electrons/positrons.  However, the sharp cutoff feature in the low energy side of the spectrum is almost independent of these properties.  In Fig.~3, we can see that the low energy cutoff of each spectrum is slightly broadened compared with that in Fig. 4.  This is because the model adopted in this
 figure assumes that $\varepsilon_{\rm esc}(t)$ decreases more rapidly than in the case of Fig.~4 and so the CR electrons/positrons in the broader energy range can reach the observer while in the case of Fig.~4 where $\varepsilon_{\rm esc}(t)$ decreases slowly the low energy cutoff becomes very narrow.  In either case the dropoff in the low energy side of the spectrum is so steep that one should assume the intrinsic spectral index as hard as $\alpha \lesssim 0-1$ if we neglect the energy-dependent CR escape effects.  {As we mentioned in the last section, if the magnetic field
 
\begin{figure}

\plotone{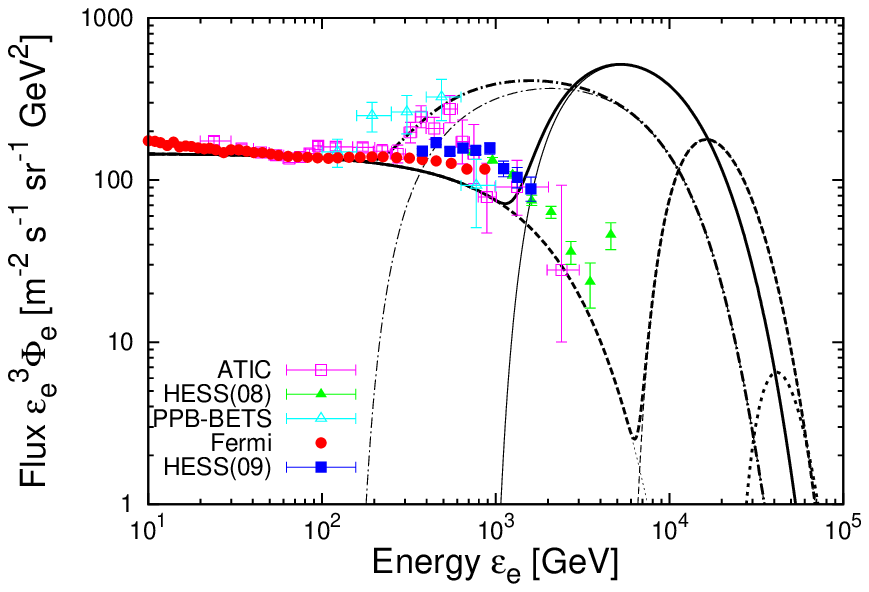}

\caption{The electron spectra from a single pulsar surrounded by a supernova remnant with different ages; $3\times 10^3$ (dotted line), $5\times 10^3$ (dashed line), $1\times 10^4$ (solid line) and $2\times 10^4{\rm years}$ (dot-dashed line).  As for the model of the escape threshold energy $\varepsilon_{\rm esc}(t)$ we adopt the results by Gabici et al. (2009; see Fig.~2).  We adopt the background model of exponentially cutoff power-law with an index of -3.0 and a cutoff at 1.5TeV, which is similar to that shown in Aharonian et al. (2008) and reproduces the data in $\sim $10GeV-1TeV well, and we show the spectrum including this background by thick lines.  We assume that a source at $r=290{\rm pc}$
from the Earth produces $e^{\pm}$ pairs with total energy $E_{e^+}=E_{e^-}=0.5\times 10^{48}{\rm erg}$
, duration $\tau_0=10^4{\rm year}$, spectral index $\alpha=2.0$ and the high energy break $\varepsilon_{_e,{\rm cut}}=10{\rm TeV}$.}

\label{f3}

\plotone{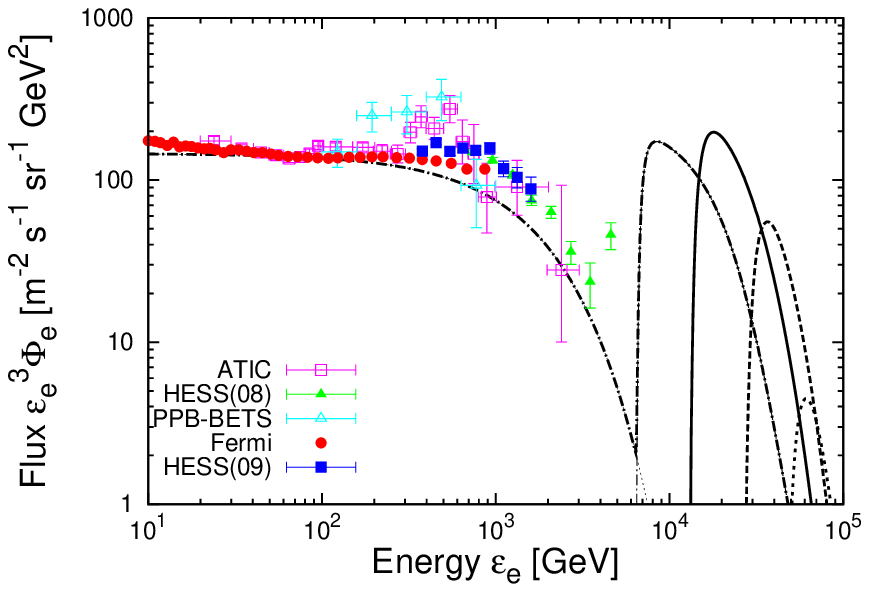}

\caption{The same plots with Fig.~3., but using the escape model by Ptuskin \& Zirakashvili (2005; see Fig. 2).}

\label{f4}

\plotone{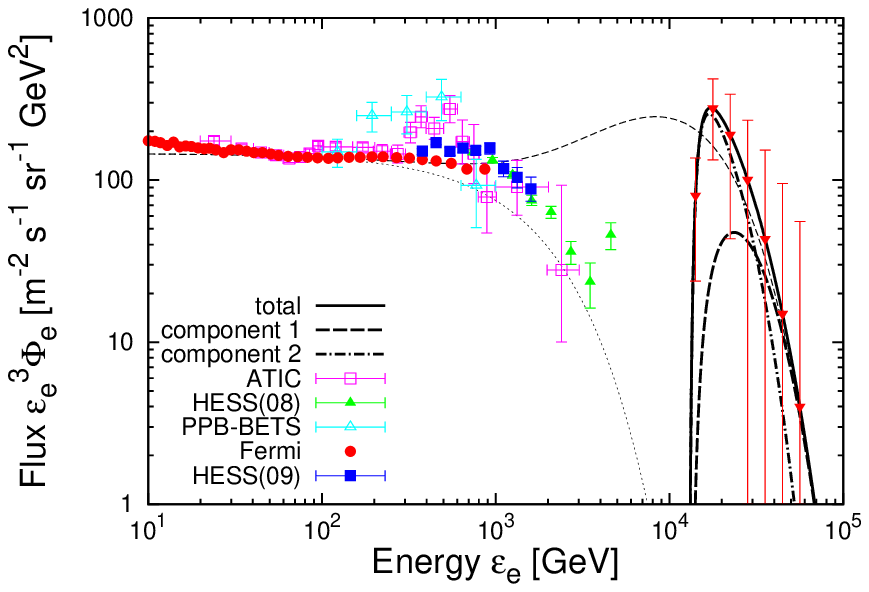}

\caption{The electron spectrum from a single pulsar surrounded by a supernova remnant (thick solid line).  As for the model for the escape threshold energy $\varepsilon_{\rm esc}(t)$ we adopt that of Ptuskin \& Zirakashvili (2005; see Fig.~2), and the background model (thin dotted line) is the same as used in Fig.~3 and Fig.~4.  The spectrum without assuming the energy-dependent escape (thin dashed line), the flux from $\dot{N}_{e,{\rm esc},1}$ and $\dot{N}_{e,{\rm esc},2}$ (long-dashed line and dot-dashed line, respectively), and the errorbars expected from the 5-years observation by CALET ($S\Omega T=220{\rm m}^2~{\rm sr}~{\rm days}$; red downward triangles) are also shown.}

\label{f5}

\end{figure}

The CR electron spectrum with the age and distance similar to the Vela pulsar ($t_{\rm age}\simeq 10^4{\rm year}$, $r\simeq 290{\rm pc}$), which is thought to be surrounded by the supernova remnant (Aschenbach et al. 1995), is shown in Fig.~5.  Here we show the spectrum
 with the escape model of Ptuskin \& Zirakashvili (2005) as well as the spectrum without the CR confinement in the SNR.  In addition, the electron flux which have not been confined in the SNR (i.e. $\dot{N}_{e,{\rm esc},1}$) and that which have once confined and escaped later ($\dot{N}_{e,{\rm esc},2}$) are shown.  We can see that the latter component dominates the flux around the low energy cutoff.  In this case we can detect the electron flux from Vela pulsar with such a sharp spectral cutoff by near future missions such as CALET (we assume the geometrical factor $S\Omega$ times the observation time $T\simeq 5{\rm years}$ as $\sim 220{\rm m}^2{\rm sr}~{\rm days}$) because the assumed electron/positron flux is sufficiently large and the low energy cutoff comes beyond the high energy dropping of the background flux inferred by H.E.S.S. (Aharonian et al. 2009).  If such a sharp cutoff feature is confirmed, that would be the strong evidence of the energy-dependent escape of CR particles from the acceleration site.  As we mentioned in the last section the second component $\dot{N}_{e,{\rm esc},2}$ would not be observed if the magnetic field in the PWN is strong and the electrons/positrons once confined in the PWN/SNR lose their energy too rapidly to escape the SNR later.  However, the first component $\dot{N}_{e,{\rm esc},1}$ also has a sharp cutoff in the low energy side of its spectrum and so even in this case it is still possible to prove the CR escape from an SNR from the electron spectrum.
If the
 threshold energy $\varepsilon_{\rm esc}(t)$ at that time is smaller than a few TeV then the cutoff feature would be
 hidden by the background flux.  In fact, assuming the phenomenological model of $\varepsilon_{\rm esc}(t)$ adopted by Gabici et al. (2009) and Ohira et al. (2010a), the low energy end of the spectrum from the Vela pulsar would be buried in the background flux because the escape energy at the age of Vela SNR ($\sim 10^4{\rm year}$) becomes smaller than $\sim $ a few TeV, and so the cutoff feature due to the energy-dependent escape of CR electrons/positrons may not be resolved.

As we mention in Sec.~1, the CR escape from a source (i.e. SNR) seems to be an important process for the observed CR spectrum below the knee ($\sim 10^{15.5}{\rm eV}$) and the broken power-law gamma-ray spectra observed from SNRs interacting with molecular clouds.  As for the spectrum of CR protons and nuclei we can see only the superposed flux from multiple CR sources with different ages, and so it is impossible to see the direct evidence of energy-dependent CR escape from a single CR source.  On the other hand, in the $\sim 1-10{\rm TeV}$ band of the CR electron spectrum we may be able to get the spectrum from a single source, and then we expect the sharp low energy cutoff showing the boundary between runaway CR electrons/positrons and confined CRs whose energy is not high enough to escape from the SNR shock.
  Therefore, this cutoff can be the first direct evidence that the energy-dependent CR escape really occurs at the source of those CR electrons/positrons.

\section{Summary and Discussions}

In this paper we show the possibility of getting the evidence of the energy-dependent CR escape from the SNR shock by observing the TeV CR electron spectrum.  The escape of CR particles is the fundamental process in emitting the CR particles accelerated at the SNR shock, and the history of the escape energy $\varepsilon_{\rm esc}(t)$, which is the threshold energy of accelerated particles not to be confined but to escape the shock into the ISM, is essential in determining the observed CR spectrum.  Although there are some indirect suggestions to this scenario from recent gamma-ray observations of SNRs interacting with molecular clouds, we have not verified the energy-dependent CR escape model of SNRs directly from the observations.

In order to see the history of CR injections from SNRs we focus on the lepton component of CRs.  In the energy band larger than $\sim {\rm TeV}$, it is expected that the electron/positron flux from a few nearby young astrophysical sources (e.g. Vela pulsar) can be observed, then we will be able to get the information about the intrinsic CR spectrum from a single source at a certain time, which has not been well understood because the observed nuclear CR spectrum consists of the contributions from multiple sources and it is impossible to resolve the spectrum from a single source.  Especially if a nearby young pulsar embedded in an SNR emits sufficiently large flux of CR electrons/positrons, they are injected into the SNR shock and only the electrons/positrons with energy larger than $\varepsilon_{\rm esc}(t)$ can go through the SNR and propagate into the ISM.  As a result, the observed electron spectrum from that pulsar may have the low energy cutoff corresponding to the escape energy $\varepsilon_{\rm esc}(t)$ at the observation time.  In order to detect such spectral feature the low energy cutoff (i.e. $\varepsilon_{\rm esc}(t)$ at the observation time) should come above $\gtrsim 1-10{\rm TeV}$ because otherwise the cutoff would be buried in the background flux and would be hard to resolve.

The possibility that there is such a low energy cutoff in TeV electron spectrum has firstly pointed out by this study, and taking into account the variation of the CR escape history from the SNR and/or the age of the pulsar, we can predict a variety of spectral shapes that have not been considered in the context of astrophysical sources.  For example, if the high energy cutoff of the spectrum, which is determined by the intrinsic cutoff energy at the pulsar and the radiative energy loss of electrons/positrons during their propagation, is nearly equal to the low energy cutoff determined by the escape from the SNR, the resulting spectrum would have a narrow-line like shape and it would be the similar feature to that expected from the annihilation of dark matter particles.  Whether the observed CR electron spectrum is originated from a single nearby source would be determined by measuring an anisotropy of CR electrons/positrons (Mao \& Shen 1972; B\"{u}esching et al. 2008; Ioka 2010).

Note that the high energy break
energy of the electron spectrum from a pulsar $\varepsilon_{e,{\rm cut}}$ generally depends on time and should be determined from the analysis of the particle acceleration processes in the pulsar wind nebula as well as the radiative cooling of electrons/positrons during the acceleration.  However, these processes have not been fully understood yet and so it is difficult to give the maximum electron/positron energy from the first principle.  For the present purpose, we are interested in the existence of the low energy cutoff due to the energy-dependent CR escape and unless the intrinsic high energy break $\varepsilon_{e,{\rm cut}}$ is much smaller than the escape energy $\varepsilon_{\rm esc}$ at the observation time, there would be a sufficiently large electron/positron flux above $\varepsilon_{\rm esc}$ and so the cutoff feature would be detected clearly enough to probe the CR escape scenario.

We should also mention that the position of the
low energy cutoff depends on the model of the
time evolution of the CR escape energy and the highly unknown magnetic and radiation fields, which produce the
energy losses of electrons/positrons.

 The energy-dependent escape of CR electrons/positrons may be confirmed by the observations of the radio to $\gamma$-ray emissions from an SNR with a pulsar.  The electrons/positrons escaping the SNR would radiate via synchrotron emissions and inverse Compton scatterings just outside the SNR shock.  If we can observe the radio to $\gamma$-ray intensity around the nearby young SNR with a pulsar and can fit that spectrum by the leptonic model with the electrons/positrons having a low energy cutoff in their energy distributions, that would be the support of the energy-dependent escape of CRs from the SNR (in preparation).
\ \\

We thank K. Kohri, A. Mizuta, K. Nakayama and Y. Suwa for useful discussions.  This work is supported in part by the World Premier International Center Initiative (WPI Program), MEXT, Japan and the Grant-in-Aid for Science Research, Japan Society for the Promotion of Science (No. 22740131 for NK; No.19047004, 21684014, 22244019, 22244030 for KI).

\end{document}